\documentclass[%
 reprint,
 amsmath,amssymb,
 aps,
]{revtex4-2}

\usepackage{subcaption}
\usepackage{graphicx}
\usepackage{bm}
\usepackage{float}
\usepackage{amsmath}
\usepackage{physics}

\begin{document}

\preprint{APS/123-QED}

\title{A trustless decentralized protocol for distributed consensus of public quantum random numbers}

\author{Lac Nguyen}
 \affiliation{Center for Quantum Science and Engineering, Stevens Institute of Technology, Hoboken, NJ 07030 U.S.A}
 \affiliation{Physics Department, Stevens Institute of Technology, Hoboken, NJ 07030 U.S.A}
 \affiliation{QPhoton, Inc. 78 John Miller Way, Kearny, NJ 07032}
\author{Jeevanandha Ramanathan}%
 \affiliation{Center for Quantum Science and Engineering, Stevens Institute of Technology, Hoboken, NJ 07030 U.S.A}
 \affiliation{Physics Department, Stevens Institute of Technology, Hoboken, NJ 07030 U.S.A}
  \affiliation{QPhoton, Inc. 78 John Miller Way, Kearny, NJ 07032}
\author{Michelle Mei Wang}%
 \affiliation{Center for Quantum Science and Engineering, Stevens Institute of Technology, Hoboken, NJ 07030 U.S.A}
 \affiliation{Physics Department, Stevens Institute of Technology, Hoboken, NJ 07030 U.S.A}
  \affiliation{QPhoton, Inc. 78 John Miller Way, Kearny, NJ 07032}
\author{Yong Meng Sua}%
 \affiliation{Center for Quantum Science and Engineering, Stevens Institute of Technology, Hoboken, NJ 07030 U.S.A}
 \affiliation{Physics Department, Stevens Institute of Technology, Hoboken, NJ 07030 U.S.A}
  \affiliation{QPhoton, Inc. 78 John Miller Way, Kearny, NJ 07032}
\author{Yuping Huang}%
 \affiliation{Center for Quantum Science and Engineering, Stevens Institute of Technology, Hoboken, NJ 07030 U.S.A}
 \affiliation{Physics Department, Stevens Institute of Technology, Hoboken, NJ 07030 U.S.A}
  \affiliation{QPhoton, Inc. 78 John Miller Way, Kearny, NJ 07032}
 \email{lnguyen1@stevens.edu}

\date{\today}

\begin{abstract}
Quantum random number (QRNG) beacons distinguish themselves from classical counterparts by providing intrinsic unpredictability originating from the fundamental laws of quantum mechanics. Most demonstrations have focused on certifiable randomness generators to guarantee the public that their genuineness is independent from imperfect implementations. These efforts however do not benefit applications where multiple distrusted users need a common set of random numbers, as they must rely on the honesty of beacon owners. In this paper, we formally introduce a design and proof-of-principle experiment of the first consensus protocol producing QRNs in a decentralized environment (dQRNG). Such protocol allows N number of participants contribute in the generation process and publicly verify numbers they collect. Security of the protocol is guaranteed given (N-1) dishonest participants. Our method is thus suited for distribute systems that requires a bias-resistant, highly secure, and public-verifiable random beacon.

\end{abstract}
\maketitle
\section*{Introduction} 
Quantum information science is fundamentally changing the way sensitive information is distributed, shared, utilized, and perceived. While quantum key distribution has been a global focus of research and development in this field, there are many other areas where quantum physics principles can be applied to solve previously prohibitive problems and create significant impacts. Among them is the generation and distribution of shared resources amongst distrusted parties over decentralized, unsecured networks.    

Suppose Alice, Bob, Dave, and Charles, who are four validators in a blockchain, need to choose one consensus leader among themselves to create the next block in a secure and fair manner, without relying on any centralized randomness source. Suppose a lottery service must prove to claimants that the process of declaring winners is purely random and not controlled by any central entity. The common ground of such scenarios is, in the absence of a trusted third party, how a group of mutually distrusted participants can agree on some genuine random choices for a public settlement among themselves \cite{WassermanProbibility}. This is the problem of constructing a random number generator in a decentralized environment (dRNG) where RNs are not only unbiased, unpredictable, tamper-resistant, but also \textit{publicly verifiable}. 
Post-election auditing, creating public parameters for cryptographic protocols, \cite{cryptoeprint:2015:366, cryptoeprint:2015:1249}, gambling and lotteries services, and private preserved messages \cite{Vuvuzela} are some applications requiring dRNGs. In recent years, with the booming of blockchain applications such as distributed computing platforms, smart contracts, and Proof-of-Stake (PoS) based consensus algorithms \cite{Vitalik, Cardano}, dRNG has significantly become more demanding. With the rapid rise in the field of quantum random number generators (QRNGs), one might wonder why not exploit the fundamentally unpredictable randomness from quantum process \cite{Gabriel2010, Shi_2016, samsonov2020vacuumbased, https://doi.org/10.1002/que2.8} \cite{Qi:10,HongPhysRevE2010}, \cite{YouPanAPL2014, Xu:16} to build a beacon for dRNG. 
In the past years, most QRNs research focused on creating either a high speed or a self-tested, device-independent generator. In the first scheme, QRNGs are built by fully characterized devices, offering up to Gbits speeds, comparatively as high as classical pseudo-random number generators (PRNGs), to accommodate numerical simulations or gaming service applications \cite{https://doi.org/10.1063/1.5078547,Kim:2021}. Meanwhile, the latter QRNGs type serves mostly cryptography purposes by delivering rather slower speeds but self-tested capability in which quantum randomness can be monitored and verified without depending on any trusted physical implementations \cite{PhysRevLett.114.150501, Avesani2018, Liu2018}. 
Nevertheless, these demonstrations benefit only individual users that have physical access to collect QRNs directly from their own devices. To solve the problem of dRNGs, a , users must remotely receive numbers broadcasted from a centralized source of quantum randomness. Because there is no way to verify if those outcomes truly originated from quantum process or whether they are protected in transit, the randomness of these numbers become subjective.  Beacon users have no choice but to trust the honesty of QRNG manufacturers, or rely on verification from a third party and security of the datatransfer process. 
These two considerable approach however do not solve the problem of dRNG. From the point of view of dRNG users, a QRNG remains a \textit{centralized source of randomness that they send request and receive digital RNs}, therefore, it only meets the technical trust of users but not the \textit{social trust}. dRNGs users still have to count on the honesty of some third party. For example, in 2018, NIST built a QRNG beacon where randomness is certified by the impossibility of superluminal signals. This brought RNG beacon to the next level where users no longer worry of adversarial activities on measurement settings, measurement devices, or randomness seeds. Nevertheless, the loophole Bell tests must be isolated from hackers and customers \cite{Bierhorst2018}, forcing public users to put trust in the beacon owners. Clearly, a crucial element is missing in current QRNG technology: not all or no RNs users participate in verifying the randomness publicly. Building on these observations, we recognize an urgent need to unlock more potential of quantum properties into solving these problems. In this paper, we propose the first dQRNG protocol that, on-demand, produces one or one set of random numbers used for decision-making on a matter involving N parties. Our protocol is motivated from both quantum and classical cryptography. We show that this novel decentralized quantum random number generator (dQRNG) protocol is:\newline
- \textit{Genuine}: no participant has control over the final outcomes.\newline
- \textit{Flexible in randomness distribution}: without post-processing, QRNs follow arbitrary probability distribution specified and agreed upon by all N parties.\newline
- \textit{Decentralized}: anyone is able to participate in the QRN generation process.\newline
- \textit{Publicly verifiable}: all N parties have the power to verify the genuineness of the outcome RN(s). \newline
- \textit{Highly tolerant}: outcomes are unpredictable if at least one party is honest among N parties. \newline
- \textit{Quantum secured}: immune to quantum computer attacks.\par
We provide a general framework of how to implement our protocol into current classical and quantum network infrastructure. Along the way, we demonstrate experimentally dQRNG with four participants and report the robustness of the protocol by showing randomness testing results, probability distribution verification, and fairness for users in QRNs generation process.
\section*{Results}
\subsection*{Previous works}
To our best knowledge, there is currently no quantum protocol for dRNG. Thus, we would like to briefly explain how classical techniques generate RNs over a decentralized network before presenting our quantum solution. Many classical consensus protocols have been developed and implemented the past years \cite{9152802, GINAR2019, cryptoeprint:2016:1067}. A popular approach is producing RNs based on a commit-reveal scheme \cite{RANDAO}. In the commit phase, each participant submits the hash of their own secret RN. In the reveal phase, all participants announce the RNs of their choices so that others can compare with the encrypted RN they sent earlier. The final RN is the outcome of a previously agreed upon function combining all secret RNs from the entire pool. A challenge this method faces is that as each participant publicizes their RNs, the last participant to reveal has an advantage of knowing the resulting RN ahead of time. Therefore, the protocol could fail because the last revealer either manipulates the output RN or refuses to submit their secret RN, thus terminating the whole RN generation process. To avoid this, RNs are fed into a verifiable delay function (VDF) with the assumption that the malicious participant does not possess specialized hardware capable of cracking the inherently slow hash computation of the VDF \cite{DanVDF}. Other directions for dRNG are built upon threshold digital signature methods where RNs are obtained by amalgamating majority of random signatures from participants \cite{NEAR}. These methods rely on public-private keys scheme and require distributed keys generation process, hence, are vulnerable against quantum computer attacks \cite{ThresholdSig}. 
\subsection*{Model}
\begin{figure}[h] 
\includegraphics[width=\linewidth]{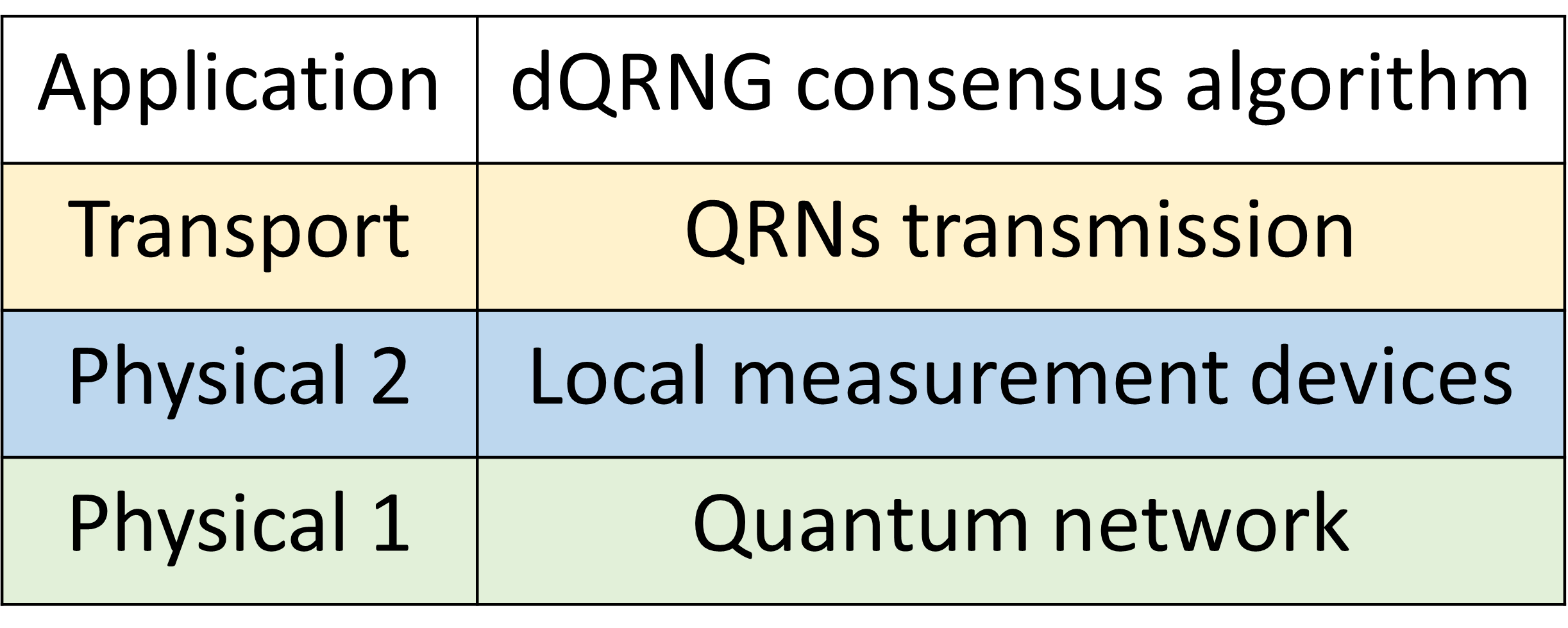}
\caption{Decentralized quantum random number generator (dQRNG) architecture}
\label{fig:dQRNGarchitecture}
\centering
\end{figure}
\begin{figure}[h] 
\includegraphics[width=\linewidth]{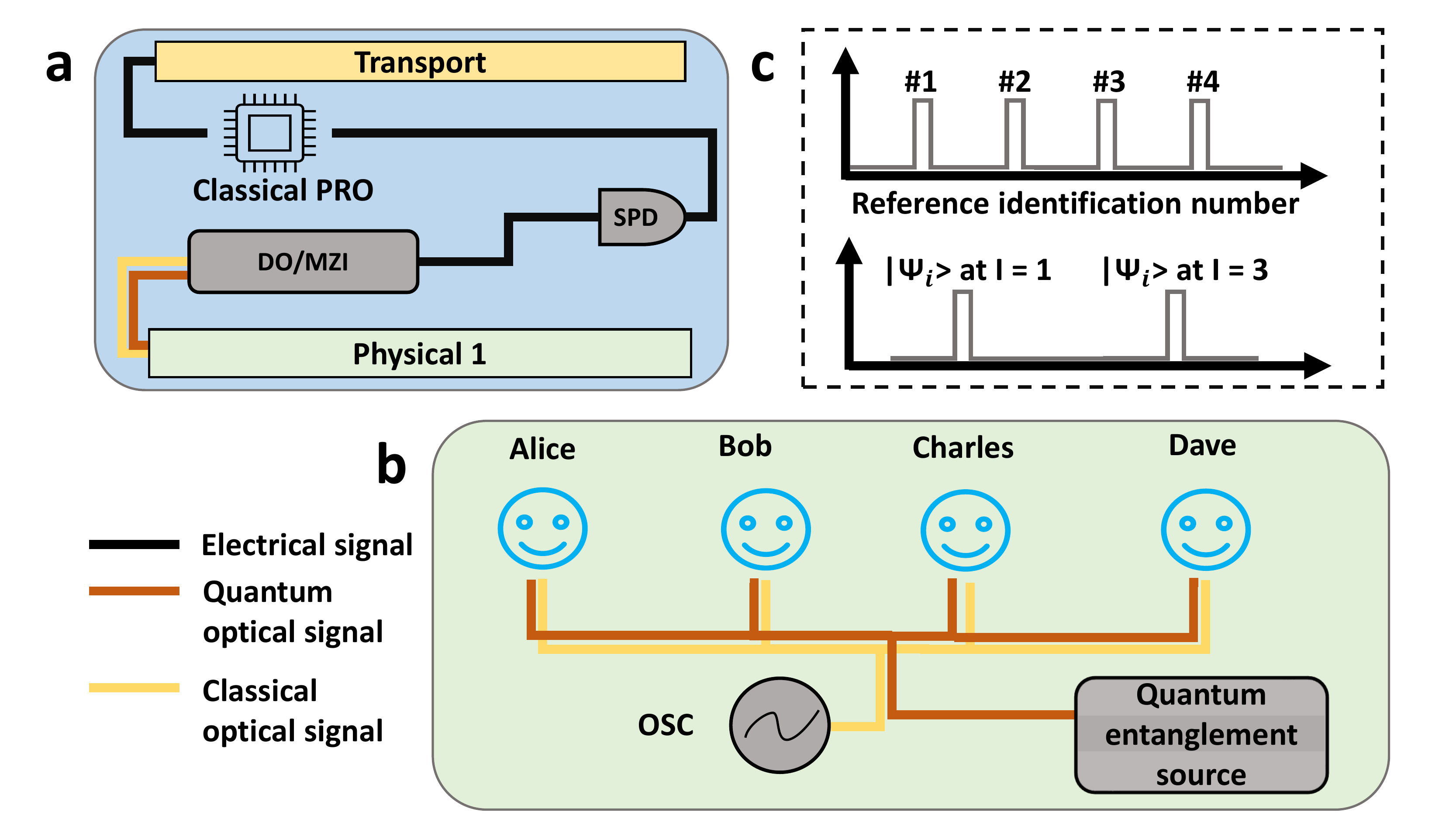}
\caption{Details of necessary devices inside physical 1 (a) and 2 (b) layers of dQRNG architecture. (c) explains the timing diagram of the synchronized reference signal for all nodes. A detected photon carries an identification (ID) corresponding to the closest reference pulse arriving before it.}
\label{fig:physicalLayers}
\centering
\end{figure}

All communication layers of dQRNG are implemented in parallel corresponding to those in classical communication \cite{Cerf1983TheDI}. Figure \ref{fig:dQRNGarchitecture} depicts quantum related layers of dQRNG. Physical layers are identical to quantum network where there must be an entanglement source, private quantum channels between communication parties, and entanglement verification setup. Detailed schematic illustration of dQRNG physical layers are described in figure \ref{fig:physicalLayers}a and \ref{fig:physicalLayers}b. 
\begin{figure*}
\centering
\includegraphics[width=\textwidth]{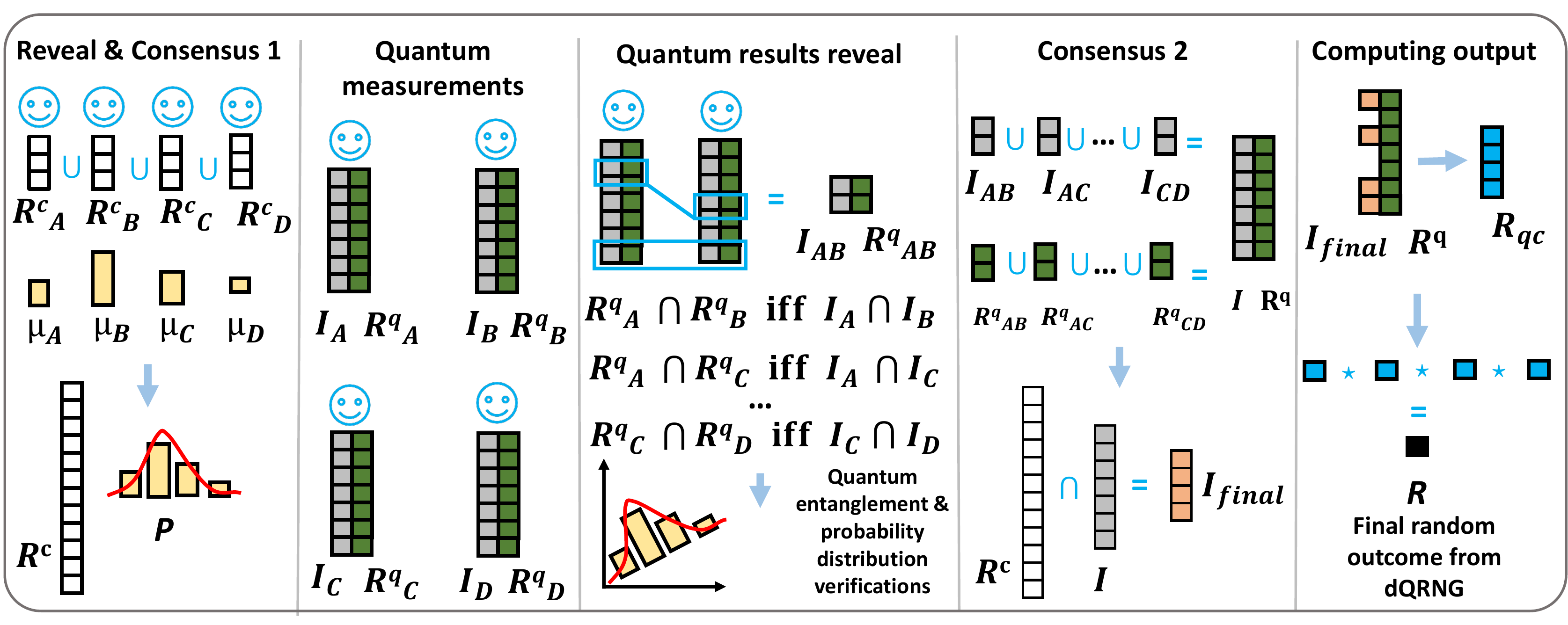}
\caption{Diagram of the dQRNG protocol procedure involving four parties, Alice, Bob, Charles, and Dave.}
\end{figure*}
We apply a model that has been reported in many previous studies as a solution for large-scale quantum networks, in which entangled photons are passed through a beamsplitter of N, distributing to N nodes such that each party shares photon pairs with every other party (total of $P_{num} = \frac{N(N - 1)}{2}$ number of possible channel pairs) \cite{doi:10.1063/5.0002595}. Each communication node must have local measurement and data processing devices such as single photon detectors and classical information processors. Transport layer in classical communication is co-used in this protocol for transmission of QRNs between nodes. Unlike quantum network architecture, our protocol can be built without the assumption of secured authentication. Consensus algorithm of dQRNG is operated on classical application layer. All parties in the communication pool are synchronized to a common reference signal of period $T$ to keep track order of their detected photon state. Such a reference can be supplied through the same optical communication channels to distribute the entangled photons (see figure \ref{fig:physicalLayers}b and \ref{fig:physicalLayers}c).
\subsection*{Construction} 

We first introduce the notations necessary for reading this protocol. Let $N$ be the number of participants, and $k$ be $(N-1)$. This pool would like to agree on a list of random numbers $R$ of length $l$ within a bounded range $B$. Depending on the application, parameter $\mu$, representing the probability of being generated of a random numbers, is decided publicly in advance \cite{Nguyen2019ProofofStakeCM,10.1093/rfs/hhaa075}. Parameter ${I}$ indicates reference identification number or pulse index representing the time slot number of a detected photon. $R\textsuperscript{c}$ is a list of classical true random numbers generated by participants' choices, $R\textsuperscript{q}$ is a list of QRNs generated by quantum process, and $R\textsuperscript{qc}$ is a list representing RNs produced from the dQRNG protocol. $F$ is a function, such as $XOR$ operation or summation operation, agreed ahead of time by participants such that $R = F(R\textsuperscript{qc})$.
We assume the existence of a quantum network infrastructure, as described in the previous section, connecting all communication nodes. All participants publicly agree upon a common photon measurement basis. We do not assume any necessary authentication between nodes nor existence of any encryption and decryption, given quantum or classical method, for data transferring during the protocol operation. A set of random numbers $R$ is generated in five steps:

\textbf{Reveal $\&$ Consensus 1}. Each participant generates and publicizes their own list of random values of length $m$ ($m >> l$) called $R\textsuperscript{c}_{i} = {\{r\textsuperscript{c}_{i,m}\}}$ and probability weight $\mu_i$ ($i$ indicates participant such as Alice (A), Bob (B), Charles(C), Dave(D), etc; this parameter is optional) $r\textsuperscript{c}_{i,m} \in N$, $r\textsuperscript{c}_{i,m} \leq B$. List of all $R\textsuperscript{c}_{i}$ are combined to create $R\textsuperscript{c}$ = $R\textsuperscript{c}_{A} \cup R\textsuperscript{c}_{B} \cup R\textsuperscript{c}_{C} \cup R\textsuperscript{c}_{D} ...\cup R\textsuperscript{c}_{N} = \{r\textsuperscript{c}_{m_c}\}$, where $m_c = m \times N$. A probability density function (pdf) $C$ is constructed by combining probability weight $\mu_i$.

\textbf{Quantum measurements}. Entangled photon pairs are sent to all participants. Each participant then performs photon measurement privately to obtain their collapsed photon states $R\textsuperscript{q}_{i} = \{\ket{r\textsuperscript{q}_{i,m\prime_{i}}}\}$ with corresponding indices $I_i = \{ind_{i,m\prime_{i}}\}$ ($m\prime$ is the number of detected photons. In practice, length $m\prime$ is different for each node depending on the loss of the quantum channel and measurement noise, $m\prime_{i} >> m >> l$).

\textbf{Quantum results reveal}. Measurement results of photon states with corresponding indices are publicly revealed by each participant. Quantum entanglement verification is performed by extracting detected photon results previous step from $P_{num}$ number of channel pairs, $R\textsuperscript{q}_{ij} = R\textsuperscript{q}_{i} \cap R\textsuperscript{q}_{j} = \{\ket{r\textsuperscript{q}_{i,m\prime_{i}}}\} \cap \{\ket{r\textsuperscript{q}_{j,m\prime_{j}}}\} = \{\ket{r\textsuperscript{q}_{ij,m\prime_{ij}}}\}$ with corresponding list of indices $I_{ij} = I_i \cap I_j$. In the ideal case, $m\prime_{ij} = m\prime_{i} = m\prime_{j}$, however, $m\prime_{ij} < m\prime_{i}$ and $m\prime_{j}$ due to quantum channel loss and measurement noise. For example, for pair Alice and Bob, correlated list of $R\textsuperscript{q}_{AB} = R\textsuperscript{q}_{A} \cap R\textsuperscript{q}_{B} = \{\ket{r\textsuperscript{q}_{A,m\prime_A}}\} \cap \{\ket{r\textsuperscript{q}_{B,m\prime_B}}\}$ with corresponding list of indices $I_{AB} = I_A \cap I_B = \{ind_{A,m\prime}\} \cap \{ind_{B,m\prime}\} = \{ind_{AB,m\prime_{AB}}\}$. \\
Similarly, using the same measurement results, probability distribution verification is performed to compare with pdf $C$.

\textbf{Consensus 2}. A list of QRNs is obtained by appending lists of correlated photon measurement results $R\textsuperscript{q}_{ij}$ from all $P_{num}$ channel pairs, forming $R\textsuperscript{q} = R\textsuperscript{q}_{AB} \cup R\textsuperscript{q}_{AC} \cup R\textsuperscript{q}_{AD} \cup R\textsuperscript{q}_{BC} ... =  \{\ket{r\textsuperscript{q}_{m \prime \prime}}\}$, for all $P_{num}$ and $I = I_{AB} \cup I_{AC} \cup I_{AD} \cup I_{BC} ...= \{ind_{m\prime \prime}\}$, for all $P_{num}$ where $m^{\prime\prime} = \sum_{P_{num}} m\prime_{ij}$.\\ Quantum randomness is combined with users' randomness by first publicly perform $I \cap R\textsuperscript{c}$ to assemble a new list of indices called $I_{final} = \{ind_{m\prime \prime}\} \cap \{r\textsuperscript{c}_{m_c}\} $. Then, the final list of random numbers is filtered so that only values with corresponding indices in $I_{final}$ remain, called $R_{qc} = \{r\textsuperscript{qc}_{m_c}\}$. It can be seen that $R_{qc}$ are selected as subset of QRNs harvested in the quantum process by consensus happened before quantum measurements were operated, and therefore, still are QRNs. 

\textbf{Computing output}. To obtain the list of $R$ random numbers of length $l$ as desired, an aggregation function $F$ (e.g modular addition) is used to calculate publicly and can be confirmed by any participants if needed, $R = F(\{r\textsuperscript{qc}_{m_c}\})$.

\subsection*{Analysis}
\begin{figure*}[t]
\includegraphics[width=\textwidth]{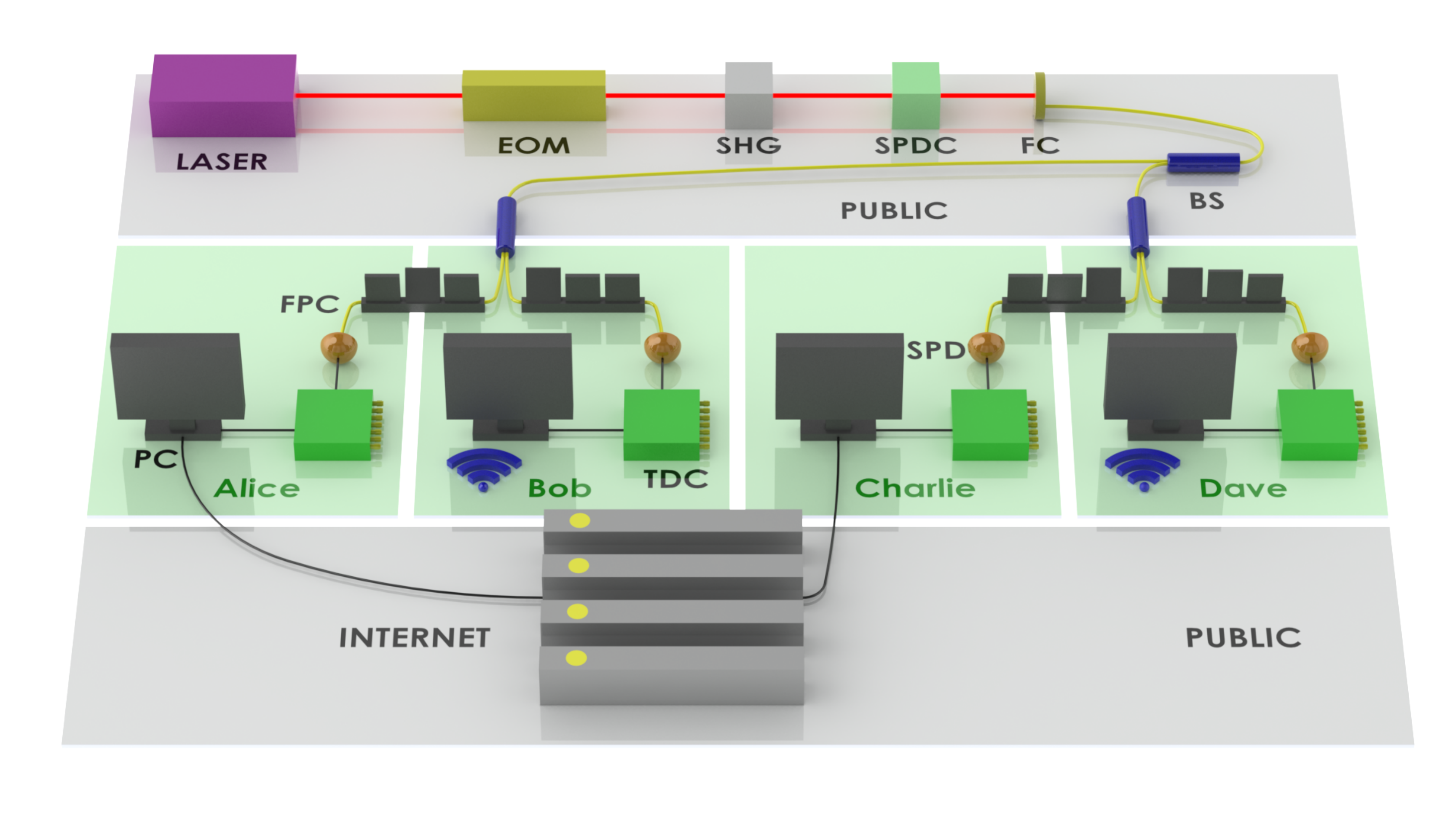}
\caption{Experimental setup to demonstrate four parties performing random number generation consensus}
\label{Fig:experimentalSetup}
\centering
\end{figure*}
\textbf{Correctness}: For as long as at most $k$ participants are malicious, meaning at least 1 participants follow the protocol honestly, the results will still be unbiased and unpredictable, the protocol will be executed and all participants will agree upon the final result when the protocol is completed. \\
\textit{Proof}. Assume the case of four participants Alice, Bob, Charlie, and Dave where they need to agree on one truly random number ($l = 1$) in the range from 0 to 6, given each of number is equally likely to be selected. Alice is the only one being honest ($k = 3$). We proceed the dQRNG protocol to show that the final result is still random and incalculable. \\
\textit{Reveal $\&$ Consensus 1}. Each participant Alice (A), Bob (B), Charles(C), Dave(D) generates and publicizes their own list of random values of length $m$, let's make $m = 10$:
\begin{align*}
    R\textsuperscript{c}_{A} = {\{r\textsuperscript{c}_{A,1\rightarrow{10}}\}},
    R\textsuperscript{c}_{B} = {\{r\textsuperscript{c}_{B,1\rightarrow{10}}\}},\\
    R\textsuperscript{c}_{C} = {\{r\textsuperscript{c}_{C,1\rightarrow{10}}\}},
    R\textsuperscript{c}_{D} = {\{r\textsuperscript{c}_{D,1\rightarrow{10}}\}},\\
    R\textsuperscript{c} = R\textsuperscript{c}_{A} \cup R\textsuperscript{c}_{B} \cup R\textsuperscript{c}_{C} \cup R\textsuperscript{c}_{D}
   = \{r\textsuperscript{c}_{1\rightarrow{40}}\}, m_c = 40\\
    C = 1
\end{align*}
\textit{Quantum measurements}. 
\begin{align*}
    R\textsuperscript{q}_{A} = {\{r\textsuperscript{q}_{A,m\prime_{A}}}\},    
    R\textsuperscript{q}_{B} = {\{r\textsuperscript{q}_{B,m\prime_{B}}}\},\\
    R\textsuperscript{q}_{C} = {\{r\textsuperscript{q}_{C,m\prime_{C}}}\},
    R\textsuperscript{q}_{D} = {\{r\textsuperscript{q}_{D,m\prime_{D}}}\},\\
     (m\prime_{i} >> 10 >> 1).
\end{align*}
\textit{Quantum results reveal}. 
Alice reveals her results honestly while Bob, Charles, and Dave do not. The three of them might cheat without knowing there are other cheaters in the pool. They could alter their results partially or entirely with artificial data. This method, however, does not favor the cheaters in anyway. Quantum entanglement verification easily displays them because their announced results must be compared with each others and with Alice.\\
In another possible scenario, all three cheaters might consult each other and assemble set of data such that they can pass the quantum entanglement verification with each other in public. At the same time, they still embedded some honest results to verify with Alice. In such situation, 
\begin{align*}
   R\textsuperscript{q}_{AB} = R\textsuperscript{q}_{A} \cap R\textsuperscript{q}_{B} (\text{non-deterministic}),\\\nonumber
   R\textsuperscript{q}_{AC} = R\textsuperscript{q}_{A} \cap R\textsuperscript{q}_{C} (\text{non-deterministic}),\\\nonumber
   R\textsuperscript{q}_{AD} = R\textsuperscript{q}_{A} \cap R\textsuperscript{q}_{D} (\text{non-deterministic}),\\\nonumber
   R\textsuperscript{q}_{BC} = R\textsuperscript{q}_{B} \cap R\textsuperscript{q}_{C} (\text{deterministic}),\\\nonumber
   R\textsuperscript{q}_{BD} = R\textsuperscript{q}_{B} \cap R\textsuperscript{q}_{D} (\text{deterministic}),\\\nonumber
   R\textsuperscript{q}_{CD} = R\textsuperscript{q}_{C} \cap R\textsuperscript{q}_{D} (\text{deterministic}),\\\nonumber
\end{align*}
with corresponding indices behave in the same manner $I_{AB}, I_{AC}, I_{AD}$ are non-deterministic; $I_{BC}, I_{BD}, I_{CD}$ are deterministic. \\
\textit{Consensus 2}. $R\textsuperscript{q} = R\textsuperscript{q}_{AB} \cup R\textsuperscript{q}_{AC} \cup R\textsuperscript{q}_{AD} \cup R\textsuperscript{q}_{BC} \cup R\textsuperscript{q}_{BD} \cup R\textsuperscript{q}_{CD}$ and therefore, it's a combination of random numbers extracted from quantum process and falsified ones made up by Bob, Charles, and Dave. Likewise, $I = I_{AB} \cup I_{AC} \cup I_{AD} \cup I_{BC} \cup I_{BD} \cup I_{CD}$ is a combination of random indices and fabricate ones. Assume in the first step, $R\textsuperscript{c}_{B},R\textsuperscript{c}_{C}, R\textsuperscript{c}_{D}$ are agreed and arranged by Bob, Charles, and Dave before they proclaim, $R\textsuperscript{c}$ still carries private randomness comes from Alice. Thus, the outcome indices of $I_{final} = I \cap R\textsuperscript{c}$ remains unpredictable. By post-selecting only numbers with indices $I_{final}$ in $R\textsuperscript{q}$ list,  $R_{qc} = \{r\textsuperscript{qc}\}$ is composed from mixture of quantum random numbers and arbitrary subset of predetermined numbers.\\
\textit{Computing output}. As a consequence, computing $R = F(\{r\textsuperscript{qc}\})$ returns an unpredictable number.

Now let's examine some common attacks on quantum networks and how dQRNG protocols are immune against them. 

\textbf{Digital emulation attack}. The quantum results reveal step occurs over the classical network, it is vulnerable to digital emulation attack. Without the access to the quantum channel, an adversary can pretend to be one of the participant because the consensus happens in the application layer of the classical channel. Observing others revealing their results first, he then fabricates $R\textsuperscript{q}_{i} = \{\ket{r\textsuperscript{q}_{i,m\prime_{i}}}\}$ with corresponding indices $I_i = \{ind_{i,m\prime_{i}}\}$ to emulate that he has matching measurement results with every other participants. This malicious action is rather meaningless because it neither helps him know the final random numbers result before hand nor manipulate them.

\textbf{Man-in-the-middle attack (MITM)}. Since we do not assume any secure quantum authentication requirement, an outside adversary can perform MITM attack and take the identity of that node. In this case, the adversary gains full access to quantum measurement results and has the choice to publicize any type of results he wants. His dishonest reveal, if he chooses to do so, does not affect the unpredictability of the final results as proved previously. Participants do not hide their quantum measurement results and corresponding indices, therefore, knowing these information ahead of time does not give the adversary any advantages either. 

\subsection*{Results}
\begin{table}[h]
    \centering
    \begin{tabular}{|c | c | c |}
    \hline 
     Case  &  1 & 2\\
     \hline \hline
     N    & 4 & 4\\
     \hline
     R    &  8-bit (binary) value & Number represents A, B, C or D\\
     \hline
     B     &  [1,256] & [0,3]\\
     \hline
     l     &    1    &  1 \\
     \hline
     $\mu_i$ & N/A & $\mu_A = \mu_B = \mu_C = \mu_D = 0.25$\\
     \hline
     $R\textsuperscript{c}_i$ & generated by PRNG & generated by PRNG\\
     \hline
     m     &   8388607     & 8388607\\ 
     \hline
     $R\textsuperscript{q}_{i}$ & collected from & collected from\\
     & quantum measurements & quantum measurements\\
    \hline
    \end{tabular}
    \caption{For this typical example, users A, B, C, and D received $m\prime_A=1029054$, $m\prime_B=964265$, $m\prime_C=952180$, and $m\prime_D=1019553$ photons respectively. This parameter will vary each time the protocol is run. Other parameters that will vary are user-input random number list $R_c$, matching pairs $R^q_{ij}$ with indices $I_{ij}$ where $i,j=A,B,C,D$}
    \label{tab:dQRNG_experimtal_parameters}
\end{table}

\begin{figure}[h]
\includegraphics[width=\linewidth]{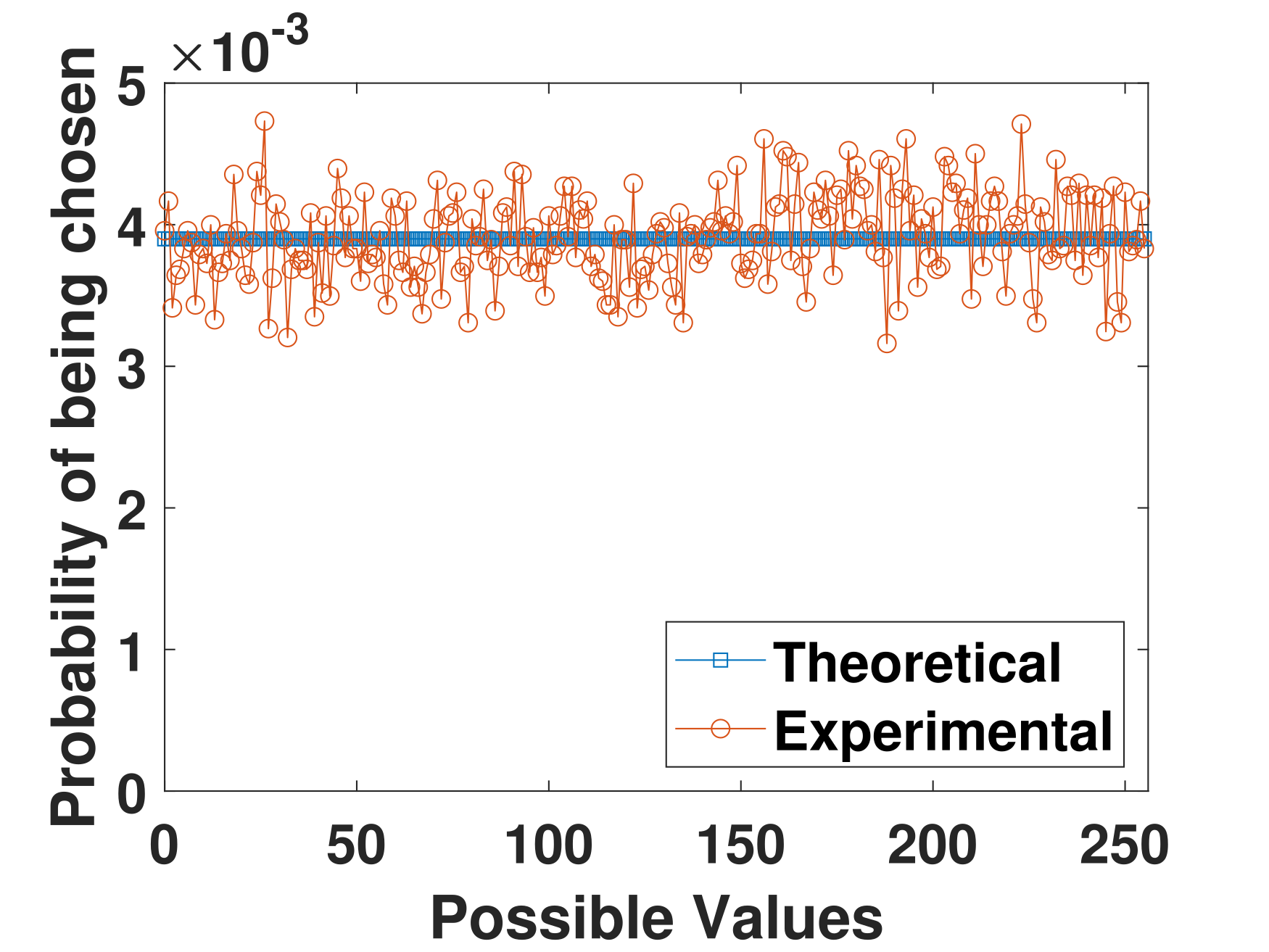}
\caption{Case 1: dQRNG protocol is used when four parties must together repeatedly generate an 8-bit random number after a certain time interval for cryptography purposes. The protocol is run 100 times with various amount of 8-bit random numbers produced each time. Typical results of the probability that an 8-bit random number (total 256 possible values) is generated by dQRNG protocol with four participants is recorded. Comparing to the theoretical probability of  $\frac{1}{256} = 0.0039$, experimental results fluctuate between 0.0032 and 0.0047 (this gives errors less than $20.5\%$).}
\label{Fig:dQRNGSimulationUniformCase1}
\centering
\end{figure}

\begin{figure}[h]
\includegraphics[width=\linewidth]{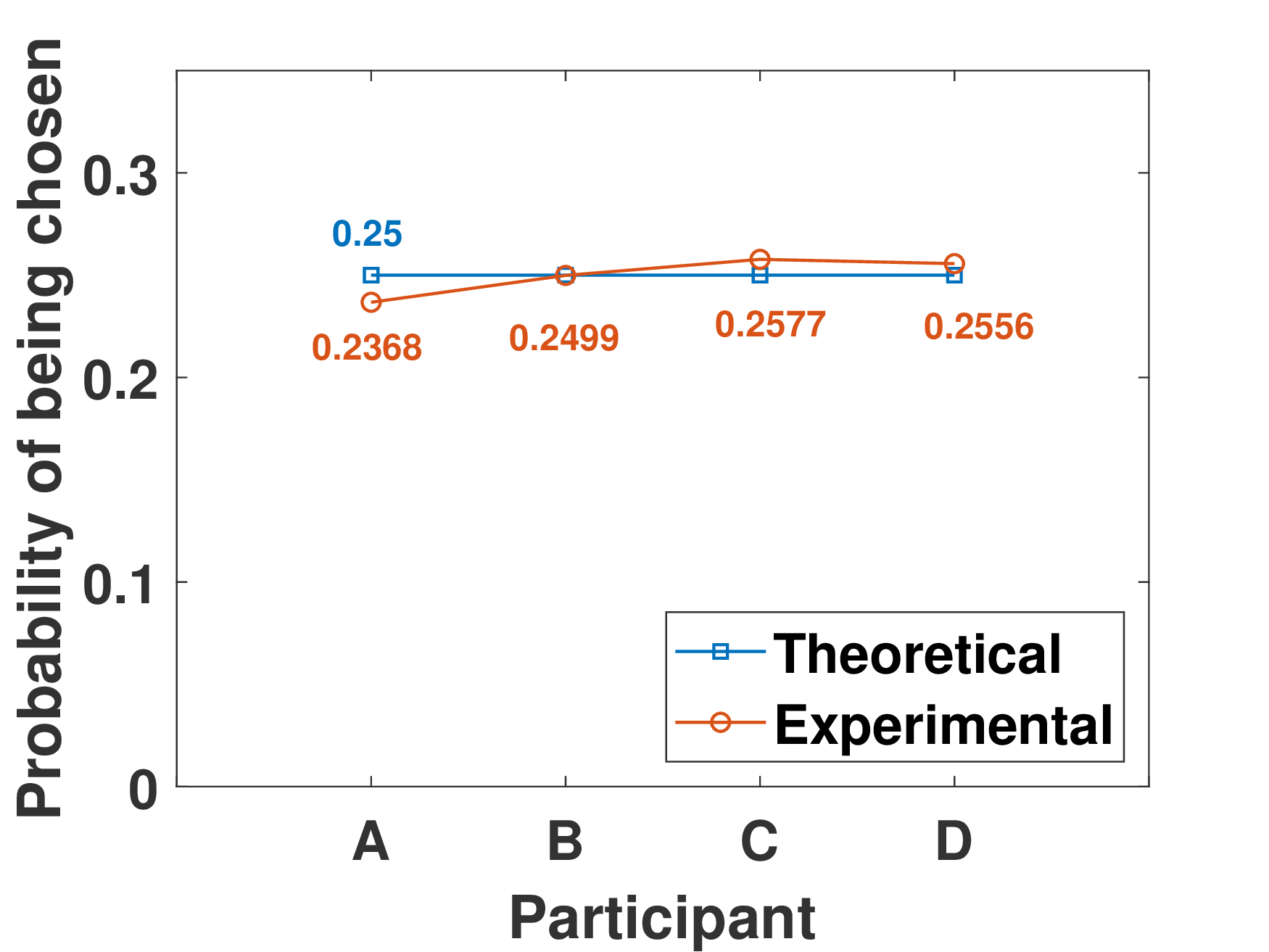}
\caption{Case 2: dQRNG protocol is used when four parties must together select a winner among themselves for some voting protocol purposes. The protocol is run 100 times with one party is chosen each time. Comparing to the theoretical probability of 0.25 experimental results only fluctuate less than $5.28 \%$ showing the protocol is unbiased.}
\label{Fig:dQRNGsimulationuniformcase2}
\centering
\end{figure}
Here we realize the dQRNG protocol in setups purposing for different applications. Table \ref{tab:dQRNG_experimtal_parameters} records parameters used and measured in each experiment. The first type of scenario occurs when a pool of parties repeatedly shares a new random string for cryptography protocols such as private messages or multi-party computation\cite{Private}. In particular, we create a case where four parties must generate an 8-bit string periodically. We use pseudo-random number generators (PRNG) to simulate choices of $R\textsuperscript{c}_i$ list from each party. The rest of the dQRNG steps are executed following the raw data collected from the experiment whose details are explained in section Methods. To examine the uniformity of the dQRNG, we run protocol 100 times with multiple random numbers string are produced each time. Figure \ref{Fig:dQRNG_simulation_uniform_case1} indicates the average  probability of generating a certain 8-bit value calculated from 100 experiment runs comparing to the theoretically probability. 
The second type of dQRNG use case happens in scenario requiring a random fair voting mechanism. Specific examples could be a winner must be decided in a lottery or a validator must be selected to create the next block in a blockchain protocol. In our experimental demonstration, a pool of four parties must decide who is chosen with the condition that each party equally likely has chance of being chosen ($\mu_i = 0.25$). Similar to case 1, PRNGs are used to generate list $R\textsuperscript{c}_i$ in step 1 of dQRNG while quantum measurement outcomes are used for rest of the steps. We run our protocol 100 times and recorded the average probability results in Figure \ref{Fig:dQRNG_simulation_uniform_case2}.

\begin{figure*}[t] 
  \centering
  \subcaptionbox{}[.3\linewidth][c]{%
    \includegraphics[width=\linewidth]{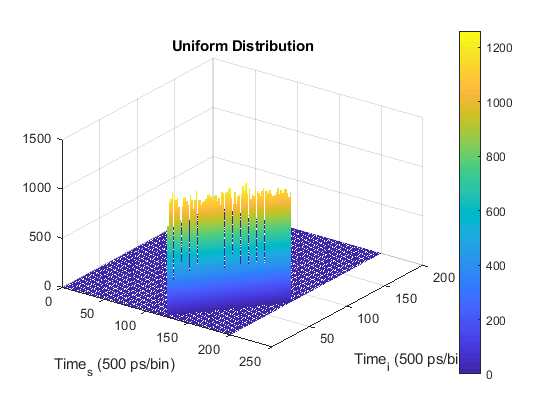}}
 \subcaptionbox{}[.3\linewidth][c]{%
    \includegraphics[width=\linewidth]{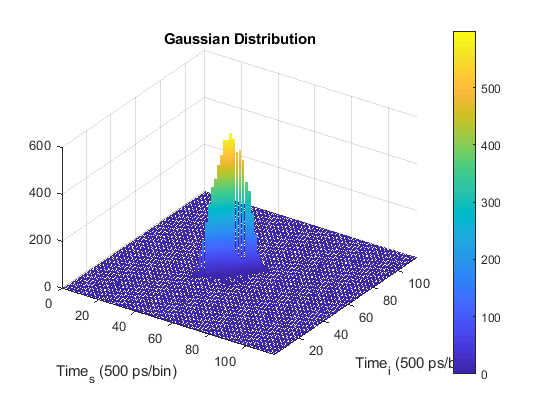}}
  \subcaptionbox{}[.3\linewidth][c]{%
    \includegraphics[width=\linewidth]{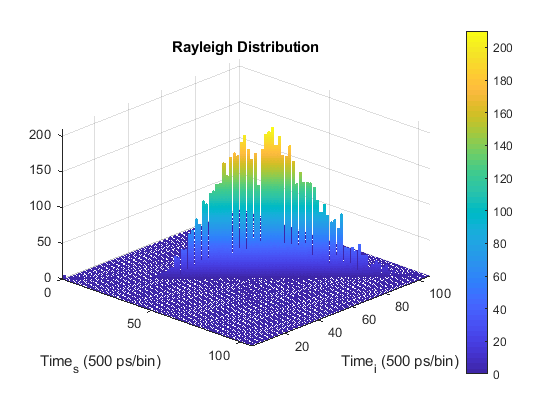}}
\caption{Consider the independent timing jitters of the SPD (33 ps) and the TDC (20 ps) at each communication node, photon arrival times collected are rounded to 250 ps per bin-width to compensate the overall timing jitter of all nodes in the network. Joint photon arrival time plots which present correlated QRNs of a typical channel pair are showed in (a), (b), and (c) following uniform, gaussian, and rayleigh probability distributions.}
\label{Fig:JTI}

\end{figure*}


\begin{figure*}
\centering
\begin{minipage}[b]{.4\textwidth}
\includegraphics[width=\linewidth]{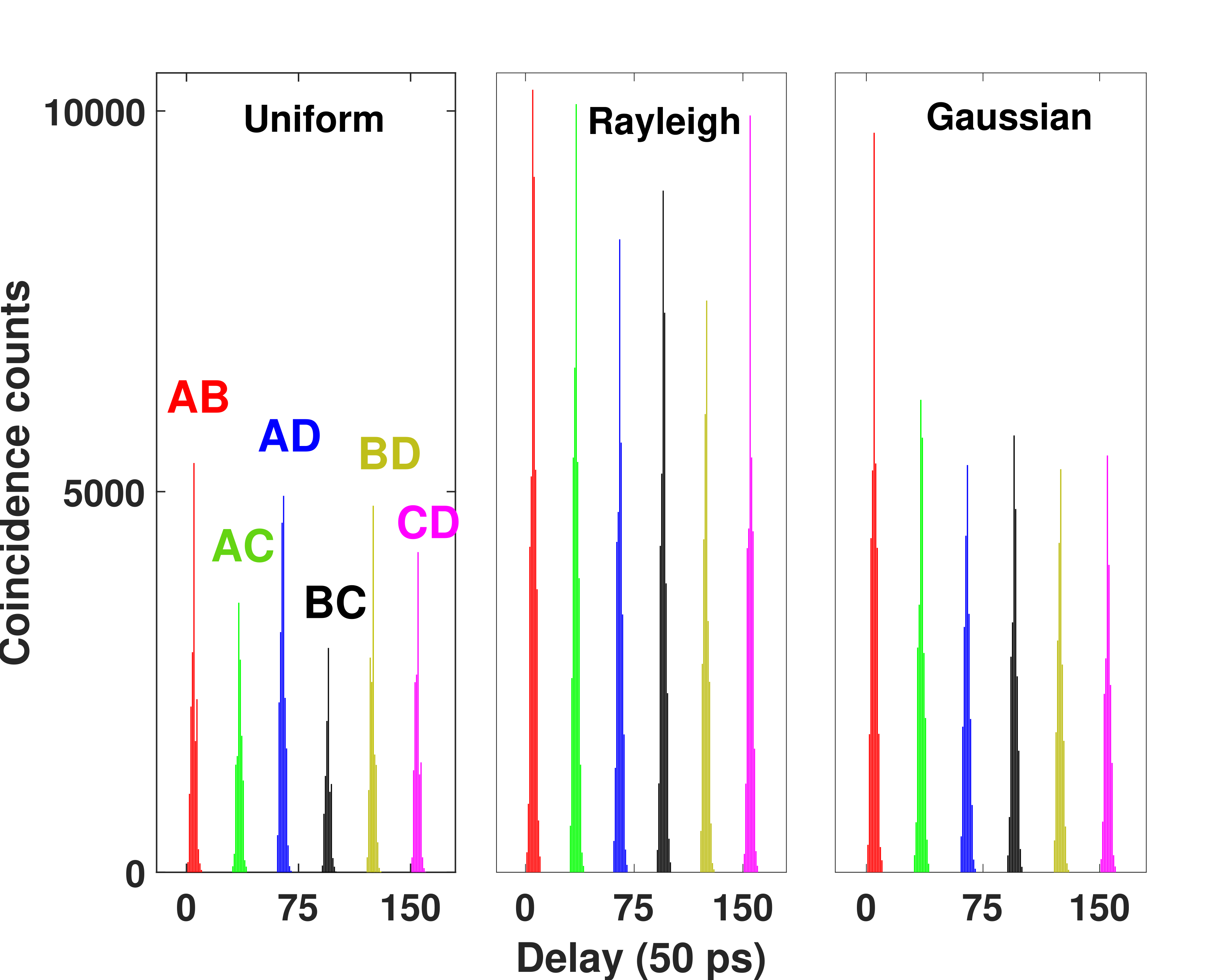}
\label{fig:Coincidence}
\end{minipage}\qquad
\begin{minipage}[b]{.4\textwidth}
\includegraphics[width=\linewidth]{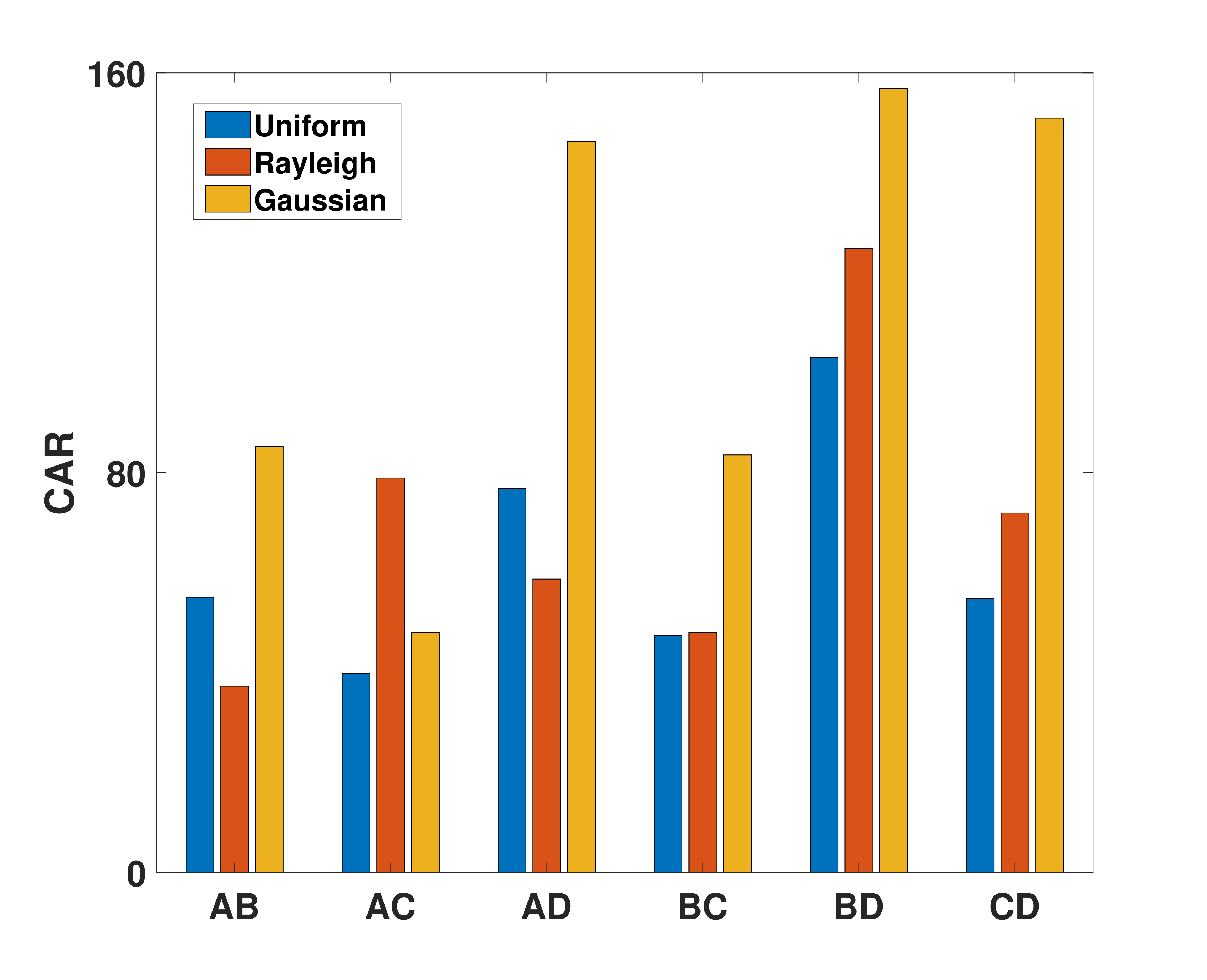}
\label{fig:CAR}
\end{minipage}
\caption{Left figure: Temporal cross-correlation between photon arrival times of each channel pair in different probability distributions. From left to right are uniform, Gaussian, and Rayleigh distributions. Total coincidence photon pairs are collected over 60s interval delaying within 500 ps around the maximum peak. Right figure: Accidental coincidence counts with photon arrival times rounded to 250 ps binwidth }
\label{fig:CAR-coin}
\end{figure*}



\section*{Discussion}
The invention of distributed applications and blockchain have promised to bring security, fairness, and transparency to digital technologies, business solutions, as well as socioeconomic and political structure by enabling consensus decision in a global scale. Nonetheless, this relies on dRNG and heavily depends on keeping it conspiracy-resistant using current classical communication techniques. It will be deleterious when a fast enough quantum computer breaks the current dRNG and from there, vandalizes this decentralized-everything idea completely. One might argue, a natural approach to quantum proof dRNG could be applying quantum cryptography protocols such as quantum keys distribution, quantum secret sharing, or quantum digital signatures to protect data transfer or signatures encryption \cite{Hillery_1999, Collins_2017}. However, in trying to make a scalable and applicable dRNG protocol, we create and present a quantum version that is simple, lightweight. Our dQRNG avoids the cumbersome procedures in quantum cryptography including keys distillation, quantum error correction, or slow key rates, but still holds the quantum advance in being provably random and information-theoretic secure. At the same time, comparing to classical domain, our protocol operates without colluding given only one honest party while requiring no encryption algorithm, and thus, reduces the communication complexity between parties. Furthermore, with the arbitrary probability distribution QRNs generation feature, our protocol outperforms others by avoiding an extra step of complex mathematics and not wasting random numbers when transforming QRNs from uniform distribution to others.

\section*{Methods} 
We realize our dQRNG protocol by constructing a simple experimental model comprised of four participants depicted in Figure \ref{Fig:experimentalSetup}. We first apply the same technique of generating QRNs with arbitrary probability distribution as we introduced in our previous work to create programmable QRN\cite{NguyenQRNG}. The difference is that the dQRNG protocol requires broadcasting QRNs to all participants, thus an entangled photons generation and distribution process must follow afterwards. The procedure is described as follow: first, intensity of an 1559.67 nm continuous wave (CW) laser is modulated using an electro-optical modulator (EOM) before coupled into a PPLN waveguide for second-harmonic generation (SHG) at 779.8 nm ; next, the SHG light is coupled into  another identical PPLN waveguide for generating entangled photon pairs via spontaneous-parametric-down-conversion (SPDC) process; and lastly, a system of three $50/50$ fiber beamsplitters are used to randomly transmit photons into four different nodes. Note that quantum entanglement verification is an essential procedure for dQRNG protocol. SPDC is ubiquitous for many entanglement-based quantum networks demonstrated thus far. At each node, photons are passed through fiber polarization controller (FPC) before detected by superconducting nanowire single photon detector (SNSPD). In this experiment, using a time-to-digital converter (TDC), we record the photon arrival times compared to a network synchronized 10 MHz reference signal thus harvest the high dimensional temporal mode of each single photon detection. Moreover, the dQRNG protocol is scalable to multiple degrees of freedom in photonics states (ie. polarization, spectral, and spatial modes ) for photon efficient generation of QRNs. Data are then transferred to a local machine which is connected to others in the classical network to perform consensus tasks in the application layer. 

\begin{figure}[H]
\includegraphics[width=\linewidth]{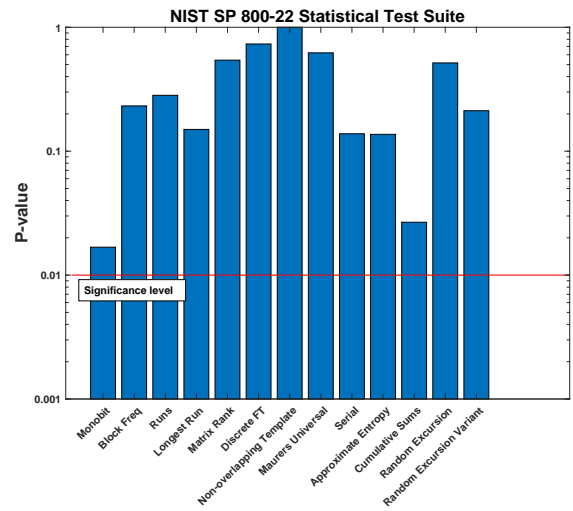}
\centering
\vspace{-2cm}
\caption{NIST results of typical 63182 correlated QRNs of uniform probability distribution. QRNs are harvested from 250 ps binwidth photon arrival times and then QRNs are converted into 8-bit representation of 505456 bits and inputted into the NIST test suite. P-values greater than significance level 0.01 certifies this sequence passes the randomness threshold rigorously for standard cryptography purposes.}
\label{NIST}
\end{figure}

In the experimental demonstration of our dQRNG protocol, three different pump shapes including CW, Gaussian, and Rayleigh pulses are sent into the SPDC source, showing the versatility in probability distribution of our QRNs. Correlated QRNs of each channel pair $R\textsuperscript{q}_{ij}$ must satisfy the designated distribution to pass the \textit{quantum results reveal} step. Figure 5 displays joint QRNs between a typical channel pair when photon arrival times are rounded to 250 ps binwidth. Figure 6 (left) describes the time resolved coincidence peaks of all available channel pairs AB, AC, AD, BC, BD, and CD with corresponding CW, Rayleigh, and Gaussian SPDC pump pulses, respectively. Discrepancies in coincidence counts between channel pairs are due to the imperfection of beamsplitters as well as contributions by different detection efficiency and timing jitters of each SPD channel. Using one SPDC source to distribute photon pairs to the entire network reduces the coincidence count rate or correlated QRNs rate. The dQRNG produces one or one small set of RNs at a time, thus, even though the average $R\textsuperscript{q}_{ij}$ rate we achieve is rather low, it does not affect the performance of the protocol. With the same time bin's width, coincidence to accidental coincidence ratios (CARs) of all possible channel pairs is described in Figure \ref{fig:CAR-coin}b. Security threshold of dQRNG protocol depends on CAR to decide when the SPDC process should end. For example, assuming CAR threshold is 50 for gaussian pump shape, photon pairs generation continue until all channel pairs achieve this value. From our experiment data, this means Alice, Bob, Charles, and Dave each has length $m\prime_A = , m\prime_B = , m\prime_C = , m\prime_D = $, respectively. It can be seen that $m\prime_i >> l = 1$ in this case. Lastly, following \textit{consensus 2} step, we append all $R\textsuperscript{q}_{ij}$ into one list $R\textsuperscript{q}$ and examine this sequence using National Institute of Standards and Technology (NIST) SP 800-22 Statistical Test Suite \cite{8986}. Results of a typical QRNs sequence in uniform distribution is presented in figure \ref{NIST}, showing our QRNs pass these test suites with high confidence level. 

\nocite{*}
\bibliography{main}

\end{document}